\documentclass[10pt,twocolumn,letterpaper]{article}

\usepackage{cvpr}              % To produce the CAMERA-READY version
\usepackage{graphicx}
\usepackage{amsmath}
\usepackage{amssymb}
\usepackage{booktabs}
\usepackage{bm}
\usepackage{tabularx}
\usepackage{multirow}
\usepackage{multicol}
\usepackage{wrapfig,booktabs}
\usepackage{algorithm}
\usepackage{algpseudocode}
\usepackage[dvipsnames]{xcolor}

\definecolor{cvprblue}{rgb}{0.21,0.49,0.74}
\usepackage[pagebackref,breaklinks,colorlinks,citecolor=cvprblue]{hyperref}

\def\paperID{16705} % *** Enter the Paper ID here
\def\confName{CVPR}
\def\confYear{2024}

\title{Laplacian-guided Entropy Model in Neural Codec with Blur-dissipated Synthesis}

\author{Atefeh Khoshkhahtinat, Ali Zafari, Piyush M. Mehta, Nasser M. Nasrabadi  \\
West Virginia University, WV, USA\\
\tt\small \{\href{mailto:ak00043@mix.wvu.edu}{ak00043},\href{mailto:az00004@mix.wvu.edu}{az00004}\}@mix.wvu.edu,\{\href{mailto:piyush.mehta@mail.wvu.edu}{piyush.mehta},\href{mailto:nasser.nasrabadi@mail.wvu.edu}{nasser.nasrabadi}\}@mail.wvu.edu}

\begin{document}
\maketitle
\begin{abstract}
  While replacing Gaussian decoders with a conditional diffusion model enhances the perceptual quality of reconstructions in neural image compression, their lack of inductive bias for image data restricts their ability to achieve state-of-the-art perceptual levels. To address this limitation, we adopt a non-isotropic diffusion model at the decoder side. This model imposes an inductive bias aimed at distinguishing between frequency contents, thereby facilitating the generation of high-quality images. Moreover, our framework is equipped with a novel entropy model that accurately models the probability distribution of latent representation by exploiting spatio-channel correlations in latent space, while accelerating the entropy decoding step. This channel-wise entropy model leverages both local and global spatial contexts within each channel chunk. The global spatial context is built upon the Transformer, which is specifically designed for image compression tasks. The designed Transformer employs a Laplacian-shaped positional encoding, the learnable  parameters of which are adaptively  adjusted for each channel cluster. Our experiments demonstrate that our proposed framework yields better perceptual quality compared to cutting-edge generative-based codecs, and the proposed entropy model contributes to notable bitrate savings.
\end{abstract}    
\section{Introduction}
Image compression is a crucial tack in image processing which aims to decrease the amount of data required for storing or transmitting without significant loss of visual content. Recently, learning-based image compression methods \cite{liu2023learned,balle2018variational,he2021checkerboard,yang2022lossy,zafari2023frequency} have demonstrated the potential to surpass classical hand-engineered codecs in terms of rate-distortion performance. Learned image compression commonly comprises three steps: transformation, quantization, and lossless entropy coding, which  resemble the components found in the traditional transform coding paradigm \cite{yang2023introduction}.
\begin{figure}
  \centering
  \scalebox{.7}{\includegraphics{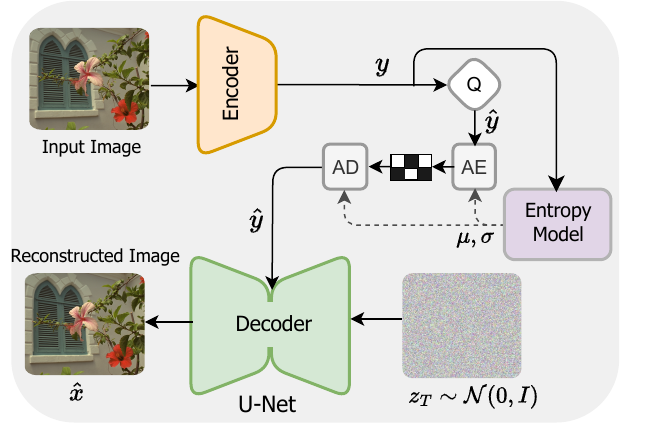}}
  \caption{Overview of our proposed neural codec. The quantized semantic latent variable $\bm{\hat{y}}$ is utilized by a diffusion-based decoder to generate realistically reconstructed image.}
  \label{fig:qplot}
\end{figure}

Neural image compression methods are typically trained with two primary objectives: minimizing distortion between the original input image and its reconstructed version and reducing the bitrate for efficient data transmission or storage \cite{theis2016lossy}. However, solely optimizing for rate-distortion may lead to blurry reconstructions. To address this issue, various generative-based codecs \cite{mentzer2020high,ghouse2023residual,yang2022lossy,agustsson2023multi} have emerged, aiming to generate reconstructions that are not only faithful to the original data but also visually realistic to human observers.\cite{mentzer2020high} introduced HiFiC, an image compression network based on GANs that aims to achieve a balance between fidelity and perceptual quality. The network demonstrated substantial advantages for human observers, surpassing BPG even when using half the bits. However, HiFiC faces challenges due to the instability of GAN training and necessitates various design adjustments. Inspired by the achievements of Denoising Diffusion Probabilistic Models (DDPMs)\cite{ho2020denoising} in generative modeling, authors in \cite{yang2022lossy} replaced the Gaussian decoder of VAEs with a conditional diffusion model to mitigate the issue of blurriness in decoded images. Despite the impressive performance of diffusion-based network, its diffusion and denoising processes do not explicitly incorporate the inductive biases inherent to natural images. The imposing inductive biases, such as the multi-scale nature of images, enable the generation of high-quality samples \cite{hoogeboom2022blurring}.

In addition to enhancing the quality of reconstructed images, accurate entropy estimation of latent representations is vital for boosting compression efficiency. The more the estimated entropy gets closer to the actual entropy of the latent, the lower the bitrate required for generating bitstream file. In this respect, several works have been proposed. Ball{\'{e} \emph{et al.}\cite{balle2018variational} introduced an entropy model that is conditioned on an additional latent variable, named the hyper-prior, to capture the existing spatial redundancies within the latent space. Studies \cite{lee2018context,minnen2018joint} introduced a sequential autoregressive block into the entropy model. This block leverages causally adjacent latent elements to estimate the distribution of the current latent, which results in slowing down of the decoding step. He \emph{et al.}\cite{he2021checkerboard} utilized a two-way parallel context model as a substitute for the sequential autoregressive context model, aiming to accelerate the decoding of latent code. In another realm of studies, to mitigate the slow decoding times, authors in \cite{minnen2020channel} introduced channel-wise context modeling as an alternative to serial spatial context. To expedite the extracting of the channel-conditional context model, He \emph{et al.} \cite{he2022elic} exploited an unevenly segmented channel-wise autoregressive model, wherein the channels are divided into varying sizes. Strengthening the entropy model through self-attention was initially proposed in \cite{qian2020learning}. This approach employed a global reference model to identify the most suitable spatial context, which doesn't have to be confined to the local scope. The idea of integrating global context into the entropy model was subsequently refined in a later work \cite{qian2021entroformer}, where a specially crafted transformer-based entropy model was employed. This entropy model is equipped with the diamond-relative position encoding to effectively model long-distance dependencies. The diamond-shaped relative positional encoding is devised based on the clip function, which relies on the pre-defined threshold to characterize an effective receptive field. This approach is effective in computing context based on the positions of latent elements; yet, it cannot provide an optimal receptive field due to its reliance on the clip function. Recent works\cite{koyuncu2022contextformer,jiang2023mlic} focused on developing adaptive context models to extract both channel and spatial correlations within the latent space while expediting the entropy decoding step. Authors in \cite{jiang2023mlic} proposed an entropy model named MEM, aiming to capture both local and global spatial correlations in each channel segment by using a parallel bidirectional context approach. 
However, despite MEM's attention to spatio-channel correlations, it appears to overlook the crucial role of positional encoding in computing long-range spatial dependencies.

To address the aforementioned challenges, we propose a conditional diffusion-based decoder in the neural compression pipeline to enhance the realism of reconstructed images. The traditional diffusion models are generalized into a conditional non-isotropic diffusion model to incorporate an inductive bias, considering the relative importance of each frequency component of images. As a result, each frequency component of an image undergoes diffusion at distinct rates, leading to the generation of decoded images in a coarse-to-fine manner. Furthermore, a novel channel-conditional autoregressive entropy model is introduced to efficiently take into account both local and global spatial dependencies within each channel chunk of the latent space. To leverage global spatial context, a Transformer block is designed which incorporates a Laplacian-shaped positional encoding within checkerboard-shaped self-attention module. The effective Laplacian-shaped receptive field for each channel chunk is dynamically determined by learning the positional encoding parameters during the optimization of the entire neural codec.

\section{Related Works}
\subsection{Learned Image Compression}

Neural image compression methods adopt a non-linear transform coding paradigm which is founded on the variational autoencoders (VAEs) \cite{goyal2001theoretical}.  Within this scheme, the encoder initially maps the input image $\bm{x}$ to a compact latent representation $\bm{y}$. Subsequently, quantization is applied to this latent representation, leading to discrete symbols that minimize the bit requirement. Afterward, entropy coding generates bit strings using an entropy model. Ultimately, the reconstructed image $\bm{\hat{x}}$ is generated by applying the decoder to the quantized latent representation $\bm{\hat{y}}$.  This framework is commonly trained with a trade-off between the rate and distortion, such as Mean Squared Error (MSE) \cite{theis2016lossy}.

Recent studies have attempted to enhance the realism of reconstructed images by optimizing neural codecs using a triple rate-distortion-perception loss \cite{mentzer2020high,yang2021perceptual,agustsson2023multi}. Blau and Michaeli \cite{blau2019rethinking} explored the essential balance between distortion and realism, demonstrating that within a given rate, improving distortion comes at the cost of decreasing the perceptual quality of a reconstructed image. In line with this premise  where perceptual quality  is measured as the difference between the image distribution and the distribution of reconstructions, Mentzer \emph{et al.} \cite{mentzer2020high} employed a conditional GAN within an autoencoder-based compression model to enhance perceptual quality. In a similar vein, \cite{he2022po} introduced a model that aimed to improve the realism by enriching the loss function with an adversarial perceptual loss using LPIPS \cite{zhang2018unreasonable}, along with a patch-based style loss \cite{gatys2016image}. Recently, Agustsson \emph{et al.} \cite{agustsson2023multi} developed a Multi-Realism model that merges the ELIC codec \cite{he2022elic} and PatchGAN \cite{chang2019free}, showing improved performance compared to HiFiC.

%Recent studies have aimed to enhance the realism of reconstructed images by optimizing neural codecs using a triple rate-distortion-perception loss. Blau and Michaeli explored the fundamental trade-off between distortion and realism, demonstrating that within a given rate, improving distortion comes at the cost of decreasing the perceptual quality of a reconstructed image. According to the definition where perceptual quality is a difference between the image distribution and the distribution of reconstructions, mentzar add a conditional Generative  adverserial network with an autoencoder based compression model. Sara proposed a model that focused on enhancing realism by augmenting the loss function with an adversarial perceptual loss using LPIPS [49], in conjunction with a patch-based style loss.

\subsection{Diffusion Models}
Diffusion models \cite{ho2020denoising,sohl2015deep}, which belong to the family of score-based generative models, acquire the data distribution through a gradual iterative denoising process, starting from a Gaussian distribution and progressing towards the actual data distribution. Recently, they have received considerable attention owing to their training stability and high quality image generation compared to GANs. Diffusion models are adopted in various domains, including image and video generation \cite{ho2022cascaded,ho2022imagen,ramesh2022hierarchical}, super-resolution \cite{saharia2022image}, inpainting \cite{lugmayr2022repaint}, deblurring \cite{whang2022deblurring}, compression \cite{theis2022lossy,yang2022lossy,ghouse2023residual} .

In the domain of image compression, several diffusion-based codecs have been proposed. The research conducted by  Theis \emph{et al.} \cite{theis2022lossy} involved employing a general unconditional diffusion model for transmitting Gaussian samples in a lossy manner. Their approach leverages a reverse channel coding concept, which stands apart from the conventional transform coding scheme. Although their methodology performs well, it suffers from high computational cost which makes it unfeasible for dealing with high-resolution images.  Yang and Mandt \cite{yang2022lossy} introduced a neural codec, called CDC, in which the decoder takes the form of a DDPM, conditioned on a quantized latent representation.  Ghouse \emph{et al.} \cite{ghouse2023residual} began by optimizing an autoencoder network using a rate-distortion loss. Subsequently, they train a conditional diffusion model on the output of the decoder, with the goal of enhancing its perceptual quality.

%Despite its innovative characteristics, it suffers from high computational cost which makes it unfeasible for dealing with high-resolution images.

%model a data distribution via iterative denoising
%synthesizing relastic images
%by gradually denoising a Gaussian variable through a matkove chain (Markovian Process).
%are capable of generating  high-quality images comparable to those generated by Gan, while not suffering from mode collapase or instabilities.

%generate samples from agiven distribution by gradually removing noise.

%that are trained to predict an image from random noise through a gradual denoising process.

%caught a surge of intersts

\subsection{Neural Entropy Model}
The main purpose of the entropy model is to estimate the joint probability distribution over the quantized latent representation. When the learned entropy model precisely matches the true distribution of the latent representation, a lower bit-rate is required to generate a compressed file. 
Entropy estimation can take advantage of two key principles: backward and forward adaptation.  Forward adaptation uses an extra latent variable, called hyper-prior \cite{balle2018variational}, as a side information to capture spatial dependencies between elements of the latent representation. Backward adaptation, on the other hand, employs an autoregressive-based context model which leverages the previously decoded elements of the latent to predict the probability of the current element. The context model of backward methods can extract relationships between the current symbol and previously decoded symbols in various dimensions, including spatial and channel axes \cite{minnen2018joint,he2021checkerboard,minnen2020channel,he2022elic,qian2020learning,qian2021entroformer,zhu2021transformer,koyuncu2022contextformer,jiang2023mlic}.

Minnen \emph{et al.}\cite{minnen2018joint} incorporated a local spatial context with a hyper-prior network to precisely predict distribution of the latent. This type of context model can exploit local correlations between the current latent and its neighboring causal elements using masked convolutions, which consequently necessitates the adoption of serial decoding. To expedite the decoding process, He  \emph{et al.}\cite{he2021checkerboard}  evenly partitioned the latent into two groups: anchors and non-anchors. The anchors are exclusively decoded using the hyper-prior, while the non-anchors utilize both the hyper-prior and the local context model. This approach allows for the parallel decoding of both anchors and non-anchors.

An alternative strategy for parallelizing the decoding process is to extract inter-channel dependencies. In \cite{minnen2020channel} the latent code is divided into evenly sized chunks along the channel dimension, with each of these chunks being decoded using information from previously decoded ones. To speed up the channel-wise context extraction, He \emph{et al.}\cite{he2022elic} partitioned channels of the latent into an uneven set of groups, allocating fewer channels to the initial groups and more channels to the subsequent ones. This uneven division  is motivated by the observation that the earlier channels possess higher entropy compared to the remaining channels.

Some works introduced adaptive context models which focus on capturing long-range spatial correlations. Qian \emph{et al.}\cite{qian2020learning} devised a global reference context model which assesses throughout previously decoded elements to detect the most similar one to the current latent. The authors in \cite{qian2021entroformer} employed a Transformer-based entropy model, called Entroformer, for capturing long-range contexts. The Entroformer benefits from diamond-relative position encoding and a top-k self-attention mechanism, both of which contribute to providing an efficient receptive field.

%This global reference is then integrated with the local context and hyper-prior model to precisely predict the distribution parameters of the current latent.
%to the improved efficiency of computing relationships.

%to boost the efficiency of computing relationships.

%to facilitate efficient computation of relationships

%He et al. [19] speed up channel-wise context extraction by partitioning channels into uneven groups, assigning fewer channels to initial chunks and more to subsequent ones. This strategy is driven by the recognition that earlier channels exhibit higher entropy than the rest.

 %This model effectively captured long-range dependencies in probability distribution estimation Recently works [18, 26, 44, 45] propose extractingglobal or long-range contexts to further boost performanceQian et al. [28] useda ViT [9] to help the entropy model capture global contextinformation. [3] proposed to look for the best spatial context, not necessar- ily the local context. Their idea was subsequently improved [4] by using the top-k most similar contexts using the self- attention mechanism.Authors in , improve the entropy Entroformer inclused top-k self-attention and a diamond-relatvie position encoding.

%to compensate for the slow decoding time

%can be sorted into two categories: spatial context and channel context.

\section{Methods}
In this work, we develop our decoder based on the blurring diffusion model \cite{hoogeboom2022blurring}, which is conditioned on the discrete latent representation and enables the differentiation among the frequency components of an image throughout the diffusion and denoising processes. The architecture of decoder will be explained in Appendix. Moreover, a novel entropy model is employed that  efficiently encodes the quantized latent representation into a binary stream.

\subsection{Blurring Diffusion Model}
We can define the blurring diffusion model as a traditional diffusion model in the frequency space, but with distinct schedules for each dimension of data \cite{hoogeboom2022blurring}. The specifics of the schedules for vectors $\bm{\alpha_t}$ and $\bm{\sigma_t}$ are crucial in this type of model, which will be discussed. 

\noindent \textbf{Diffusion Process:} The diffusion process \cite{ho2020denoising}  progressively degrades the clean image, transforming it into nearly pure Gaussian noise over the course of T time steps. Each step of the diffusion process in frequency space can be formulated as:
\begin{equation}
q(\bm{f_{t}}|\bm{f_{x}})=
\mathcal{N} (\bm{f_{t}}; \bm{\alpha_t} \bm{f_{x}},\bm{{\sigma_t}^2}\bm{I}), 
\label{eq10}
\end{equation}
where $\bm{f_t}$ and $\bm{f_x}$ are defined as $\bm{V^T z_t}$ and $\bm{V^T x}$, respectively. In such a model, each frequency component could undergo diffusion at varying rates, with the modulation of these rates being determined by the values present in the vectors $\bm{\alpha_t}$ and {$\bm{\sigma_t}$.

\noindent \textbf{Denoising Process:} The true denoising process in the frequency space, which is equal to the deblurring operation in the time domain, can be expressed as:
\begin{equation}
    q(\bm{f_{t-1}}|\bm{f_{t}}, \bm{f_{x}})=
\mathcal{N} (\bm{f_{t-1}}; \bm{\mu_{{t}\to{t-1}}} ,\bm{{{\sigma^2}_{{t}\to{t-1}}}}\bm{I}),
\label{eq11}
\end{equation}
where $\bm{{\sigma_{t \to t-1}}= \sigma_{t|t-1}\sigma_{t-1}/\sigma_t}$ and $\bm{\mu_{{t}\to{t-1}}}= (\bm{\alpha_{t|t-1}\sigma_{t-1}^2}/\bm{\sigma_{t}^2})\bm{f_t}+(\bm{\alpha_{t-1}\sigma_{t|t-1}^2}/\bm{\sigma_{t}^2})\bm{f_x}$.The actual denoising process can be approximated using a learned denoising distribution to generate data. Thus, by considering the epsilon reparametrization of Eq. \ref{eq11}, we can derive the expression for the learned denoising process in frequency space:
\begin{multline}
  p_\theta(\bm{f_{t-1}}|\bm{f_{t}}) = \mathcal{N} (\bm{f_{t-1}}; \bm{\mu_{{t}\to{t-1}}(\hat{f_x},f_t)},\bm{{{\sigma^2}_{{t}\to{t-1}}}I})\\ \text{Re-parametrization}: 
{\bm{\hat{f_x}}}=({1}/{{\bm{\alpha_{t}}}})(\bm{f_t}-{{\bm{\sigma_{t}}}} {\bm{\hat{f_\epsilon}}}).
\end{multline}
By substituting the approximation of ${\bm{\hat{f_x}}}$ into the true denoising distribution, the mean of the denoising network can be derived by $(\bm{\alpha_{t|t-1}\sigma_{t-1}^2}/\bm{\sigma_{t}^2})\bm{f_t}+(\bm{\alpha_{t-1}\sigma_{t|t-1}^2}/\bm{\sigma_{t}^2} \bm{\alpha_t})(\bm{f_t}-{{\bm{\sigma_{t}}}} {\bm{\hat{f_\epsilon}}})$.\\

\noindent \textbf{Optimization}: According to \cite{ho2020denoising}, the loss function is simplified as bellow:
\begin{equation}
      \bm{E}_{t,\bm{x},\bm{\epsilon}}[||{\bm{{f_\epsilon}}}-\bm{\hat{f_\epsilon}}||^2]=\bm{E}[||\bm{V^T}(\bm{\epsilon}-\bm{\hat{\epsilon}})||^2] \approx\bm{E}_{t,\bm{x},\bm{\epsilon}}[||\bm{\epsilon}-\bm{\hat \epsilon}||^2],
\end{equation}
where $\bm{\hat\epsilon}= \phi_\theta(\bm{z_t},t)$ and $\bm{z_t}=\bm{{V}(\bm{\alpha_t}V^Tx+\sigma_t V^T \epsilon)}$. The $\bm{\hat\epsilon}= \phi_\theta(\bm{z_t},t)$ signifies that the neural network $\phi_\theta$ takes $\bm{z_t}$ as input and predicts $\bm{\hat{\epsilon}}$. This formulation satisfies the requirement for neural networks to perform effectively in a standard pixel space. After prediction, it is sufficient to transition back and forth between frequency space and pixel space using the DCT matrix $\bm{V^T}$ and its inverse $\bm{V}$.

%\bm{\mu_{{t}\to{t-1}}}=\frac{\bm{\alpha_{t|s}}{\bm{\sigma_{s}^2}}}{\bm{\sigma_t^2}}\bm{z_t}+\frac{\bm{\alpha_{s}}\bm{\sigma_{t|s}^2}}{\bm{\sigma_t^2}}, \bm{\sigma_{t|s}}=\frac{\bm{\sigma_{t|s}}\bm{\sigma_{t-1}}}{\sigma_{t}}

%Where $\bm{\hat\epsilon}= \phi(\bm{z_t},\theta)$ and $\bm{z_t}=\bm{{V}(\bm{\alpha_t}V^Tx_t+\sigma_t V^T \epsilon_t)}$. $\bm{\hat\epsilon}= \phi(\bm{z_t},\theta)$  expresses that the neural network takes $\bm{z_t}$ as input and predicts $\bm{\hat{\epsilon}}$. This formulation can fulfill the requirement that neural networks need to perform effectively in a standard pixel space. After prediction, it is sufficient to transition back and forth between frequency space and pixel space using the DCT matrix $\bm{V^T}$ and its inverse $\bm{V}$.

\subsection{Schedules of Blurring Diffusion Model }

The schedules for the blurring diffusion model, denoted as $\bm{\alpha_t}$ and $\bm{\sigma_t}$, are obtained by combining a standard Gaussian noise diffusion schedule (with scalar values $\sigma_t$ and $\alpha_t$) along with a blurring schedule $\bm{d_t}$. Each element of the vector $\bm{\sigma_t}$ shares the same value, as identical noise is added to all frequency components. Therefore, it becomes adequate to present a schedule for a scalar value $\sigma_t$ \cite{hoogeboom2022blurring}. The noise schedule is chosen based on a variance-preserving cosine \cite{nichol2021improved}, specifically ${\sigma_t}^2=1-{\alpha_t}^2$, where $\alpha_t= \cos(t\pi/2T)$ for $t \in [0,T]$. Following \cite{rissanen2022generative}, the blurring schedule is then defined as:
\begin{equation}
    \sigma_{B,t}=\sigma_{B,max} sin(t\pi/2T)^2,
\end{equation}
where $\sigma_{B,max}$ represents a hyperparameter equal to the maximum level of blur applied to the image. This schedule, in turn, corresponds to the dissipation time through $\tau_t=\sigma_{B,t}^2/2$.
Based on the formulation discussed in the Appendix, the blurring schedule $\bm{d_t}$, which is employed for $\bm{\alpha_t}$, is defined as follows:
\begin{equation}
    \bm{d_t}=exp(-\bm{\lambda}\tau_t),
\end{equation}
where $\bm{\lambda}$ represents the vector containing squared frequencies, and $\tau_t$ corresponds to the dissipation time. To achieve a more gradual amplification of high frequencies during the denoising process, the blurring schedule $\bm{d_t}$ is adjusted to $(1-d_{min})\exp(-\bm{\lambda}\tau_t)+d_{min}$, where $d_{min}$ is set to $0.001$. Finally, by combining the Gaussian noise schedule with the blurring schedule, resultant schedule is as follows:
\begin{equation}
    \bm{\alpha_t}=\alpha_t.\bm{d_t}, \bm{\sigma_t}=\bm{1}\sigma_t.
\end{equation}
%where $\bm{1}$ is a vector of ones.
\subsection{Blurring Diffusion Model for Compression}
The rate-distortion objective in end-to-end compression resembles the loss function of a $\beta$-VAE, where a hyperparameter $\lambda$ is utilized to balance the trade-off between the bit-rate $(R)$ and distortion $(D)$:
\begin{equation}
\text{L}= D+\lambda R= {\bm{E}_{\bm{\tilde{y}}}}[-\log p_{\bm{x}|{\bm{\tilde{y}}}}(\bm{x}|{\bm{\tilde{y}}})-\lambda \log p_{\bm{\tilde{y}}}({\bm{\tilde{y}}})]. 
\label{eq17}
\end{equation}
While many neural codecs commonly employ Gaussian or Laplacian decoders, we introduce a novel approach using a conditional blurring diffusion model as the decoder. This approach aims to yield new distortion metrics that deviate from those based on Mean Squared Error (MSE) or Mean Absolute Error (MAE). In this framework, our proposed neural codec leverages two distinct types of latent variables: a \textbf{semantic} latent variable $\bm{y}$ and \textbf{texture} latent variables $\bm{z_{1:T}}$. The semantic latent variable captures and encodes the overall content and meaning of the image. In contrast, the texture latent variables are tailored to carry finer details and intricate patterns that might not be fully represented by the semantic variable alone. Notably, unlike the semantic latent variable, the texture latent variables are not compressed but synthesized during the decoding phase.

%Many neural codecs commonly utilize Gaussian or Laplacian decoders. In contrast, we present a conditional blurring diffusion model as the decoder, with the objective of achieving new distortion metric that deviate from those of MSE or MAE.In this manner, the proposed neural incorporates two types of latent variables: a \textbf{semantic} latent variable $\bm{y}$ and \textbf{texture} latent variables $\bm{z_{1:T}}$. The semantic latent variable is responsible for capturing and encoding the image's overall content and meaning. On the other hand, the texture latent variables are designed to carry finer details and intricate patterns that may not be fully captured by the semantic variable alone. In contrast to the semantic latent variable, the texture latent variables are not stored and synthesized at decoding phase.

 The forward and backward processes of the conditional blurring diffusion model can be expressed as:

\begin{align}
\begin{split}
 & q(\bm{f_t}|\bm{f_{t-1}}) =\mathcal{N} (\bm{{f_t}}; \bm{\alpha_{t|t-1}} \bm{f_{t-1}}, \bm{{\sigma^2_{t|t-1}}\bm{I}}),\\
& p_{\theta}(\bm{f_x,f_{1:T}}|\bm{\tilde y}) = p(\bm{f_T}) p_{\theta}(\bm{f_x}|\bm{f_1},\bm{\tilde y})\prod_{t=2}^{T} p_{\theta}(\bm{f_{t-1}}|\bm{f_{t}},\bm{\tilde y}) ,\\  &p(\bm{f_{T}})= \mathcal{N} (\bm{f_T};\bm{0}, \bm{I}) \\
& p_{\theta}(\bm{f_{t-1}}|\bm{f_{t}},\bm{\tilde y}) =\mathcal{N} (\bm{f_{t-1}}; \mu_{\theta}(\bm{f_{t},\tilde y},{t}),\bm{{{\sigma^2}_{{t}\to{t-1}}}I})
\end{split}
\end{align}
where $\bm{f_t}=\bm{V^T}\bm{z_t}$ and $\bm{f_x}=\bm{V^T}\bm{x}=\bm{V^T}\bm{z_o}$. 
As shown in Eq. \ref{eq17}, distortion is equivalent to the negative marginal likelihood of input data $-\log p_{\bm{x}|{\bm{\tilde{y}}}}(\bm{x}|{\bm{\tilde{y}}})$, and minimizing it is analogous to minimizing the negative marginal likelihood of the frequency representation of the data $p_{\bm{f_x}|{\bm{\tilde{y}}}}(\bm{f_x}|{\bm{\tilde{y}}})=\int p(\bm{f_x},\bm{f_{1:T}}|{\bm{\tilde{y}}}) d\bm{{f_{1:T}}}$. Since computing the marginal likelihood is intractable, we employ its ELBO with the specified diffusion and de-blurring distributions. Following this substitution, the rate-distortion objective, through the application of Jensen's inequality, can be formulated as follows:
\begin{align}
\begin{split}
& {\bm{E}_{\bm{\tilde{y}}}}[-\log p_{\bm{x}|{\bm{\tilde{y}}}}(\bm{x}|{\bm{\tilde{y}}})-\lambda \log p_{\bm{\tilde{y}}}({\bm{\tilde{y}}})]\\\leq
&{\bm{E}_{\bm{\tilde{y}}}[-{\text{ELBO}-\lambda \log p_{\bm{\tilde{y}}}({\bm{\tilde{y}}}}})],
\end{split}
\end{align}
where $\text{ELBO}={\bm{E}_q}[\frac{p_{\theta}(\bm{f_x,f_{1:T}}|\bm{\tilde y})}{q(\bm{f_t}|\bm{f_{t-1}})}]$. As explained in Appendix, the optimizing of ELBO can be simplified to:
\begin{equation}
\text{ELBO}\approx\bm{E}_{t,\bm{x},\bm{\epsilon},\bm{\tilde{y}}}[||\bm{\epsilon}-\phi_\theta(\bm{z_t},t,\bm{\tilde y})||^2],   
\end{equation}
where $\bm{z_t}=\bm{V}\bm{\alpha_t}\bm{V^T}\bm{x}+\bm{\sigma_t}\bm{\epsilon}$. Similar to \cite{mentzer2020high}, an LPIPS loss \cite{zhang2018unreasonable}is incorporated into the rate-distortion loss to enhance the perceptual quality of the reconstructed image. As the initial image at any time step can be decoded as a function of the texture latent variable $\bm{z_t}$, the semantic latent variable $\bm{y}$, and the time step $t$, i.e., $\bm{\hat{x}_t}=\bm{V}({1}/{{\bm{\alpha_{t}}}})(\bm{V^T z_t}-{{\bm{\sigma_{t}}}} {\bm{V^T\hat{\epsilon}}})$, the total loss becomes as:
\begin{align}
\begin{split}
& {\bm{E}_{t,\bm{x},\bm{\epsilon},\bm{\tilde{y}}}}[(1-\beta)||\bm{\epsilon}-\phi_\theta(\bm{z_t},t,\bm{\tilde y})||^2 \\
&-\lambda \log p_{\bm{\tilde{y}}}(\bm{\tilde{y}})
 +\beta d_{\text{LPIPS}}(\bm{x},\bm{\hat{x}_t})],
\end{split}
\end{align}
\noindent where the $\lambda$ and $\beta$ represent hyperparameters that control the trade-off between rate, distortion, and perception.

\noindent\textbf{Decoding Process:} After the training, we utilize entropy decoding on $\bm{\hat {y}}$ through the entropy model which estimates the distribution $p_{\bm{{\hat{y}}}}(\bm{\hat{y}})$. We then employ ancestral sampling to conditionally decode the image $\bm{x}$, which is equal to $\bm{z_0}$, and starts from pure Gaussian noise $\bm{z_T}{\sim}\mathcal{N}(\bm{0},\bm{I})$:
\begin{equation}
    \bm{z_{t-1}}\leftarrow \bm{V}(\bm{\hat{\mu}_{t \to t-1}}+\bm{\sigma_{t \to t-1}}\bm{V^T}\phi_\theta(\bm{z_t},t,\bm{\hat{y}}))
\end{equation}
where $\bm{\hat{\mu}_{t \to t-1}}=\frac {\bm{\alpha_{t|t-1}\sigma_{t-1}^2}}{\bm{\sigma_{t}^2}} \bm{V^T z_t}+\frac{\bm{\sigma_{t|t-1}^2}}{\bm{\alpha_{t|{t-1}}}{\bm{\sigma_t}^2}}(\bm{V^T z_t}-{{\bm{\sigma_{t}}}}\bm{V^T}\phi_\theta(\bm{z_t},t,\bm{\hat{y}}))$. The latent variables $\bm{z_{1:T}}$ are not stored but produced during the decoding.
\begin{figure}
  \centering
  \scalebox{.43}{\includegraphics{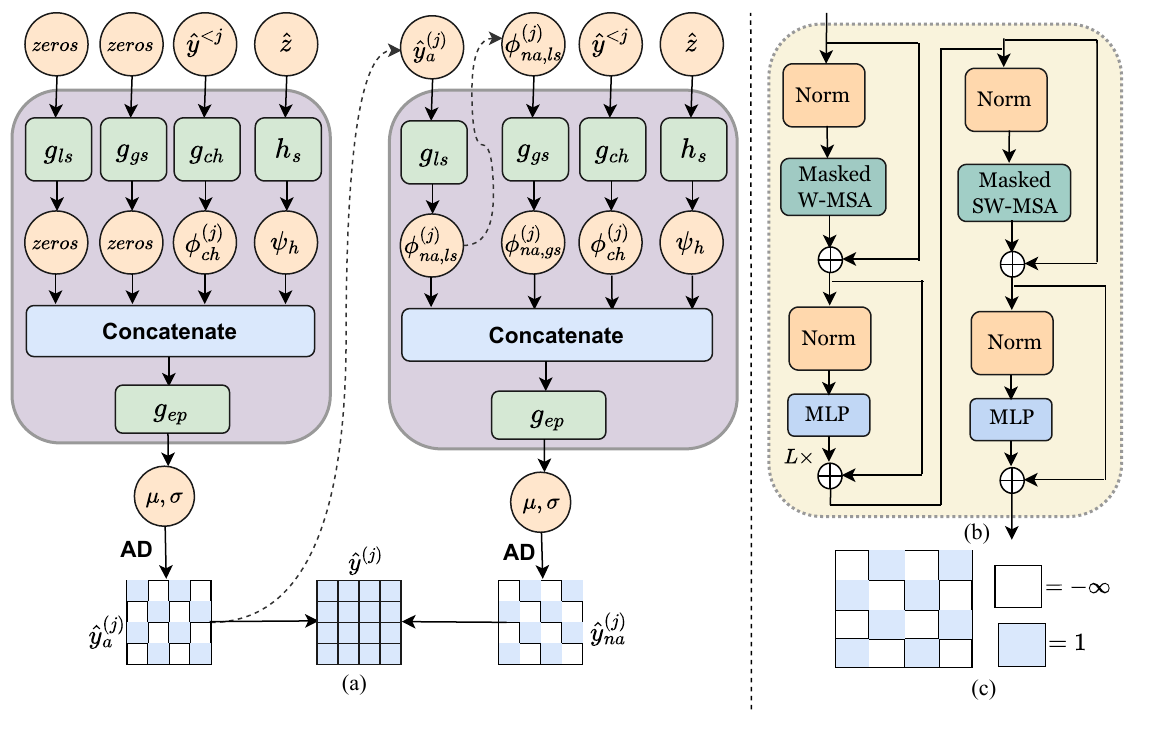}}
  \caption{ Diagram illustrating the application of the proposed entropy model for decoding the $j$-th chunk $\bm{{\hat{y}^{(j)}}}$. (b) Global Spatial Context Block. (c) An example of a checkerboard-shaped mask. }
  \label{fig:entropymodel}
\end{figure}

\subsection{Proposed Entropy Model}
The main goal of the proposed entropy model' context is to exploit both channel-wise and spatial correlations, while  expediting the decoding process. Inspired by the ELIC \cite{he2022elic}, we adopt an uneven grouping of latent channels, allowing most low entropy channels to depend on high entropy channels. The latent representation $\bm{\hat{y}}$, with $M$ channels, is clustered into five chunks along the channel dimension: 16, 16, 32, 64, and M - 128 channels, respectively. In this setup, each chunk depends on all its previous decoded chunks.  

%its previous decoded chunks
The Fig. \ref{fig:entropymodel}(a) illustrates our proposed entropy model as applied in the $j$-th chunk. Within each chunk, we use spatial context in conjunction with channel context to model correlations along both the channel and spatial dimensions. To expedite the decoding step, we employ a parallel bidirectional spatial context model which is capable of capturing both local and global spatial relationships. So, the anchor part is decoded in parallel by using solely the hyperprior and channel context, while the decoding of the non-anchor part relies on the hyperprior, as well as both the spatial and channel contexts. 
%In the following section, we will present a more comprehensive description of the spatial context.

\subsubsection{Spatial Context}

The spatial context design captures both local and global spatial correlations within the latent representation $\bm{\hat{y}}$. Following the decoding of the anchor group, a checkerboard-shaped convolution is applied to this group, generating local context for all non-anchor elements in a parallel manner. Furthermore, the acquired local contexts of the non-anchor part are subsequently fed into a Transformer-based block to efficiently extract the global spatial context. This Transformer block takes advantage of positional encoding that is customized specifically for the compression task, along with a checkerboard-based attention mechanism.
\subsubsection{Transformer-based Spatial Context}
We adopt the Swin Transformer \cite{liu2021swin} blocks, including a masked window-based multi-head self-attention (W-MSA) and a masked shifted-window-based multi-head self-attention (SW-MSA), as our global spatial context model which shown in Fig. \ref{fig:entropymodel}.(b). This selection enables us to strike a balance between computational efficiency and modeling capacity. The checkerboard-based self-attention in W-MAS and SW-MSA can be expressed:
\begin{equation}
 Atten(\bm{Q,K,V})=softmax({\bm{Q} \bm{{K}^T}}\odot\bm{M})\bm{V},   
\end{equation}where $\bm{Q} ,\bm{K}, \bm{V} \in {R}^{
 N^2 \times d }$ are the query, key, and value matrices, respectively. $N^2$ denotes the number of patches in a window  and $d$ is the dimension of query/key/value. $\bm{M}\in {R}^{N^2 \times N^2}$ represents a checkerboard-shaped mask.
 \begin{figure}
  \centering
  \scalebox{.56}{\includegraphics{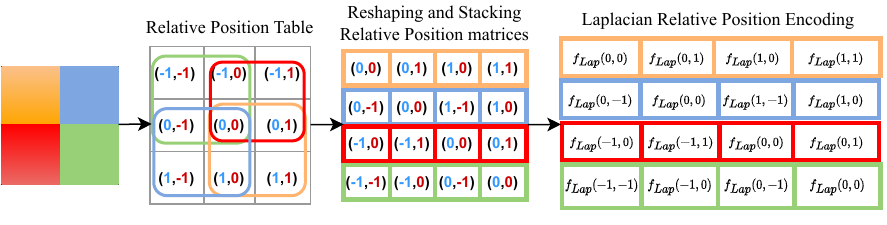}}
  \caption{ The procedure for acquiring Laplacian relative position encoding for a window with a size of $2\times 2$.}
  \label{fig:relative_position}
\end{figure}
\subsubsection{Relative Positional Encoding} The authors in \cite{he2021checkerboard,qian2021entroformer} have shown that the bitrate is impacted by the the distance of neighboring latents within the spatial context modeling of currant latent. Based on this insight, we introduce a receptive field-aware self-attention mechanism that employs a learnable Laplacian-shaped positional encoding for calculating spatial context. In our proposed approach, each channel chunk possesses its own receptive field during self-attention computation, which dynamically adjusts in response to changes in entropy of chuncks. In the $j$-th chunk, the checkerboard-based attention of W-MSA and SW-MSA is refined to receptive field-aware self-attention as follows:
\begin{equation}
    Atten(\bm{Q_j}, \bm{K_j}, \bm{V_j} )=softmax(\bm{Q_j} \bm{{k_j}^T}\odot\bm{M}+\bm{P_{Lap,j}})\bm{V_j}, 
\end{equation}
where $\bm{P_{Lap,j}}$ represents the learnable Lapalacian relative position encoding for chunk $j$. To create the Laplacian relative position encoding for the $j$-th chunk, which comprises spatial windows with a size of $N\times N$ (i.e. including $N^2$ patches), three steps need to be taken. Firstly, We generate a 2D relative position table, where each coordinate of the relative position lies in the range $[-N+1,N-1]$. Subsequently, we derive the relative position matrix for each patch. As shown in Fig. \ref{fig:relative_position}, the first patch's relative distance coordinate of (0, 0) (relative distance with its position) is located at the top-left of the orange box, while the last patch's relative distance of (0, 0) is positioned at the bottom-right of the green box. Afterward, each relative position matrix is flattened and stacked together. Finally, we apply the 2D Laplacian function to each element of the resulting matrix to generate a Laplacian relative position encoding $\bm{P_{Lap,j}}$  with learnable parameters $A_j$ and $\sigma_j$, whose size is $R^{N^2 \times N^2}$. The 2D Laplacian function is defined as:
\begin{equation}
f_{Lap}(x,y)={A_j}^2 exp((-1/2{\sigma_j}^2)(|x|+|y|)),
\end{equation}
\noindent where $x\in\{-N+1,...,N-1\}$, $y\in\{-N+1,...,N-1\}$. $A_j$ and $\sigma_j$ are learnable parameters which are determined for each chunk through optimization. The value of them is
associated with the wideness of the effective receptive field.

\begin{figure*}[tp]
    \centering
    \includegraphics[width=0.91\linewidth]{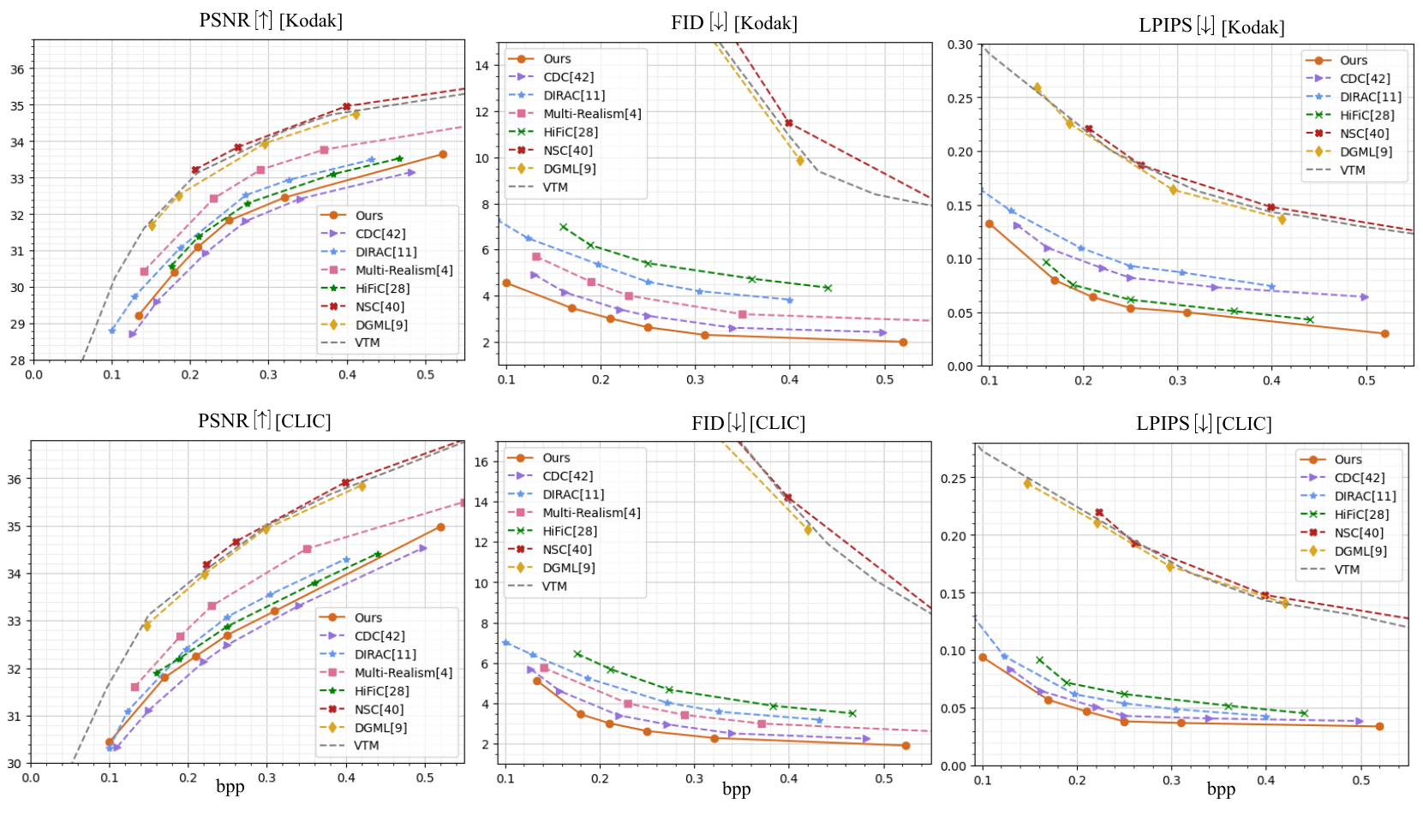}
    \caption{ Comparison of our method with other  codecs in terms of rate/distortion [bpp $\downarrow$ / PSNR $\uparrow$] and rate-perception,  including [bpp $\downarrow$ / FID $\downarrow$] and [bpp $\downarrow$ / LPIPS $\downarrow$], for both the CLIC2020 test set and the Kodak dataset. }
    \label{fig:curve-comp}
\end{figure*}

\section{Experiments}

\subsection{Implementation Details}

\begin{figure*}[tp]
    \centering
    \includegraphics[width=1\linewidth]{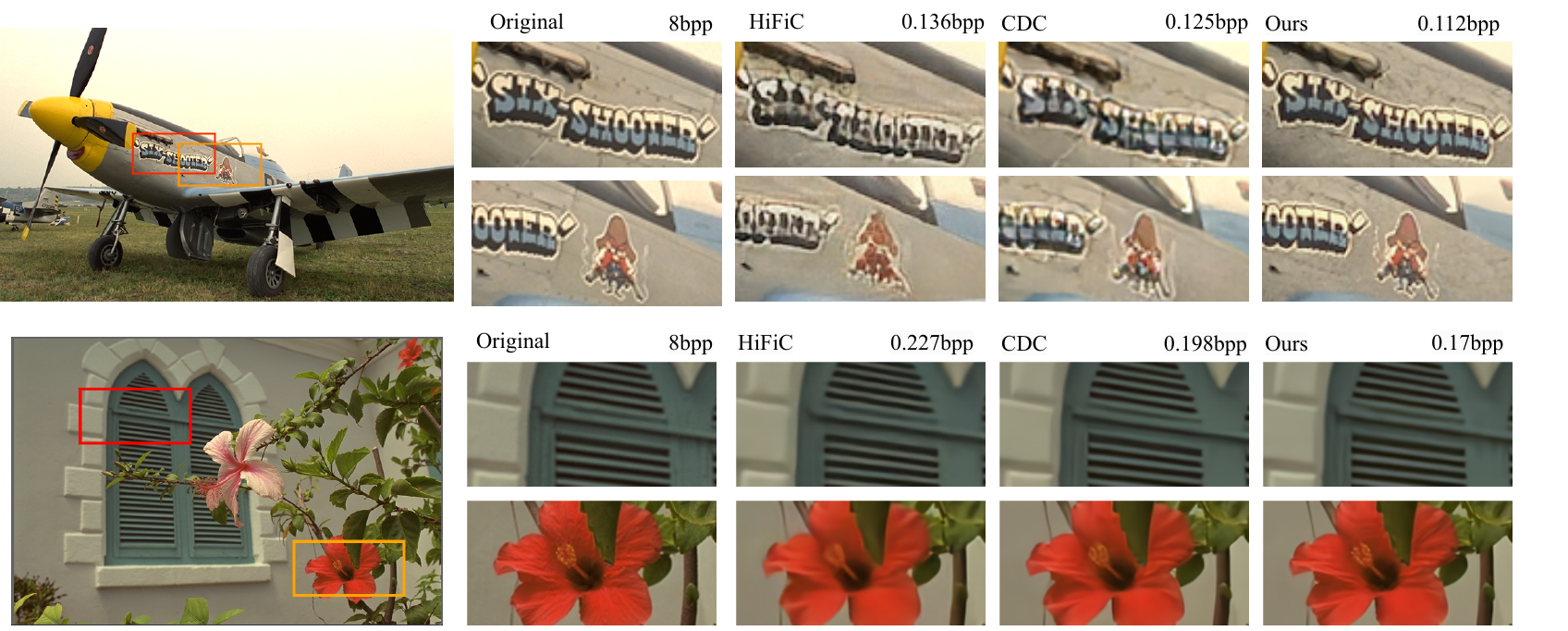}
    \caption{ Visual comparison of our method to HiFiC and CDC models shows that our model achieves superior reconstruction quality, particularly at lower bit-rates. In addition, our model displays fewer artifacts compared to both the HiFiC and CDC models.}
    \label{fig:visual-comparison}
\end{figure*}

To train our learned image compression network, we utilize a merged dataset that includes high-resolution images from the DIV2K, Flickr2K, and CLIC \cite{CLIC} datasets. These images are randomly cropped to a size of $256 \times 256$ during the training phase. 
The model parameters of all architectures were optimized using the Adam optimizer for a total of $2.4$ million steps, with a batch size of $8$. The initial learning rate was configured to be $1 \times 10^{-4}$ and was progressively reduced until the conclusion of training, reaching $1 \times 10^{-7}$. In order to cover a broad spectrum of bitrates, we selected a hyperparameter $\lambda$ from the set $\{0.0004, 0.005, 0.01, 0.02, 0.04, 0.016\}$. The hyperparameter $\beta$ which specifies the contribution of perceptual loss is considered 0.9 for all models.The parameter $T$, represents the required time steps for diffusion-based decoder, is consistently set to $500$. To evaluate the performance of our compression approach,
we select the Kodak dataset \cite{Kodak}, which comprises 24 high-quality images with a resolution of 768 × 512, and the CLIC2020 test set (428 images) with varying resolutions. 

\subsection{Comparison with the SOTA Methods}

We compare our model with state-of-the-art generative based image compression networks,  including both GAN-based and diffusion-based codecs, as well as a state-of-the-art hand-crafted codec such as VVC-Intra (VTM) \cite{vtm2022}. Additionally, we choose two VAE  codecs for comparison: DGML \cite{cheng2020learned} and NSC \cite{wang2022neural}. Specifically, within the domain of GAN-based image compression approaches, we select HiFiC \cite{mentzer2020high} and Multi-Realism \cite{agustsson2023multi} frameworks for comparison. Furthermore, we assess our method against diffusion-based models, namely DIRAC \cite{ghouse2023residual} and CDC \cite{yang2022lossy}. This comparison is performed with respect to both rate-distortion and rate-perception tradeoff. The distortion is measured by the Peak Signal-to-Noise Ratio (PSNR) metric and perceptual quality are quantified via Frechet Inception Distance (FID) \cite{heusel2017gans} and Learned Perceptual Image Patch Similarity (LPIPS). As shown in Fig. \ref{fig:curve-comp}, our model shows superior performance compared to all other codecs in terms of rate-perception. However, this heightened realism comes at the cost of distortion where other methods, except for the CDC model, either exhibit better performance or remain competitive.

\subsection{Visual Quality}
Fig. \ref{fig:visual-comparison} illustrates  reconstructed images (kodim20.png, kodim07.png) generated by our proposed model, as well as the CDC \cite{yang2022lossy} and HiFiC \cite{mentzer2020high} frameworks. Perceptually-oriented neural codecs, such as HiFiC and CDC, are criticized  when their decoding process encounters semantic content, such as text within the compressed image. Our model effectively addresses this challenge, as depicted in Fig. \ref{fig:visual-comparison}. It enhances perceptual quality through a diffusion-based decoder and imposes a strict inductive bias via the de-blurring diffusion process, resulting in both high perceptual quality and better text-preserving reconstructions. Additional qualitative comparisons are provided in the Appendix.

%Fig. \ref{fig:visual-comparison} is an instance of a reconstructed image (kodim19.png) generated by our proposed model, along with the CDC \cite{yang2022lossy} and HiFiCi \cite{mentzer2020high} frameworks. The visual comparison shows that our model tends to generate fewer artifacts and is capable of decoding image with greater realism compared to both the HiFiC and CDC networks.

%Parameters of all models were optimized using Adam by 2.4M steps with a batch size of 8. We set T = 500 for all experiments. The learning rate initial value was set to $1 \times 10^{-4}$ and gradually decreased until the end of training to $ \times 10^{-7}$. To cover a wide range of bitrates, we set the hyper-parameter  to take values from the set 0.0004, 0.005, 0.01, 0.02, 0.04, 0.016.

\subsection{Ablation study}
\noindent \textbf{Maximum Blurring:} To explore the impact of the blurring schedule, we vary the values of $\sigma_{B,max}$ and compare the performance of the resulting models. It is evident that the model with $\sigma_{B,max=0}$ is equivalent to a standard denoising diffusion model. As indicated in Table \ref{Ablationa}, blurring diffusion models with higher maximum blur levels $\sigma_{B,max}=25$ generate higher-quality reconstructed images compared to other variant models in terms of FID score.

\begin{table}[h]
  \caption{Ablation Study: all models are optimized with $\lambda$ = 0.01. (a) Investigating the impact of maximum blurring $\sigma_{B,max}$. (b) Exploring the effects of different types of positional encoding }
    \begin{subtable}[h]{0.14\textwidth}
        \scalebox{0.85}{
        \begin{tabular}{c c} % centered columns (4 columns)
\hline\hline %inserts double horizontal lines
$\sigma_{B,max}$ & FID \\ [0.2ex] % inserts table
%heading
\hline % inserts single horizontal line
0 & 3.94   \\ % inserting body of the table
5 & 3.78   \\
15& 3.56 \\
\textbf{25}& \textbf{3.45} 
 \\ % [1ex] adds vertical space
\hline %inserts single line
       \end{tabular}}
       \caption{}
       \label{Ablationa}
    \end{subtable}
    \hfill
    \begin{subtable}[h]{0.28\textwidth}
        \scalebox{0.85}{
        \begin{tabular}{c c} % centered columns (4 columns)
\hline\hline %inserts double horizontal lines
Position Enc. & bpp  \\ [0.2ex] % inserts table
%heading
\hline % inserts single horizontal line
-- & 0.2811   \\
Relative Pos. & 0.2643 \\ % inserting body of the table
Diamond Relative Pos.  & 0.2512   \\
Laplacian-shaped Pos.
 & \textbf{0.2347}  
 \\ % [1ex] adds vertical space
\hline %inserts single line
        \end{tabular}}
        \caption{}
        \label{Ablationb}
     \label{Ablation}   
     \end{subtable}

\end{table}\noindent\textbf{Positional Encoding:} We investigate the influence of various types of positional encoding in the Transformer-based entropy model. As shown in Table \ref{Ablation}, the 2D diamond relative positional encoding, which is implemented using a clip function, demonstrates better performance than the relative position encoding. However, adopting the Laplacian-shaped positional encoding results in even more significant bitrate savings compared to the 2D diamond relative positional encoding.
As depicted in Fig. \ref{fig:Receptive}, distinct values of $A$ and $\sigma$ are obtained for each channel chunk. As we progress towards the final chunks, the receptive field becomes narrower. As observed in \cite{he2022elic}, the last chunk of channels, conditioned on all previous chunks, exhibits lower entropy. Their entropy can be estimated by considering only a small neighborhood, corresponding to a narrower receptive field in our proposed relative positioning. Conversely, to provide a reasonable estimate of entropy of
the first channel chunks, a wider receptive field gathers
more information over a large context to compared to
mentioned last chunks.
%As we progress towards the final chunks, the receptive field becomes narrower. As was previously observed in \cite{he2022elic}, the last chunk of channels which are conditioned on all previouschunks, exhibit lower entropy and their their entropycan be estimated by only consideringa small neighborhood which corresponds to narrowerreceptive field in our proposed relative positioning. On theother hand, to provide a reasonable estimate of entropy of the first channel chunks, a wider receptive field gathers more information over a large context to compared to mentioned last chunks.\\

\noindent\textbf{Analysis of Context Blocks:} We conducted a comparison of the inference latency for entropy parameters during entropy decoding, as well as the bitrate savings of our proposed entropy model compared to other backward adaptation-based entropy models. For a fair comparison, all the models are equipped with a same encoder comprising of ResNet blocks and convolution layers to transform the input image $\bm{x}\in {R}^{H\times C \times 3}$ to a latent representation $\bm{y}\in {R}^{H/16\times W/16 \times 256}$. As reported in Tabel \ref{contextmodel}, our entropy model's speed is notably improved by employing unevenly grouped channels and implementing a bidirectional context model, as compared to the sequential spatial context modeling. Moreover, our findings clearly indicate that incorporating spatial global context results in superior performance when contrasted with utilizing only local context. The evaluation emphasizes that the integration of both global and local context enhances the precision of capturing spatial correlations, ultimately leading to a more accurate entropy model. In addition, our results verify that incorporating Laplacian-shaped positional encoding enhances the compression efficiency compared to MEM, which does not consider positional encoding in computing global spatial context. 

\begin{figure}
  \centering
  \scalebox{.4}{\includegraphics{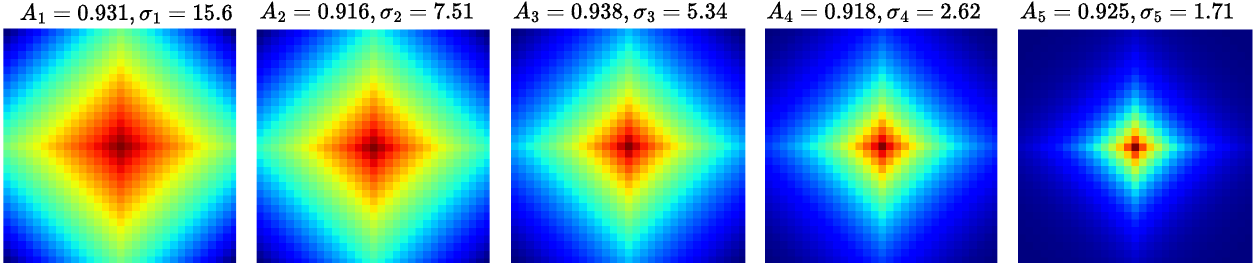}}
  \caption{ Receptive fields are extracted for each channel chunk.}
  \label{fig:Receptive}
\end{figure}

\begin{table}
\caption{The Bjøntegaard-delta-rate (BD-rate) \cite{bjontegaard2001calculation} and inference latency of entropy parameter estimation during entropy decoding (Dec.) for different context-based models on the Kodak dataset using a GPU (RTX A6000). The BD-Rate is computed relative to VVC. ([P]:parallel, [S]:Serial)}

\scalebox{0.7}{
\begin{tabular}{c|cc|c|c}
\hline\hline

\multirow{2}{*} {Method} & \multicolumn{2}{c|}{Context Model}                                                  &  {\multirow{2}{*}{Dec.(ms)} }  &\multirow{2}{*}{BD-Rate}  \\ \cline{2-3} 
                        & Channel            & Spatial &\multicolumn{1}{c|}{}  &                      \\ \hline
Minnen \emph{et. al.}\cite{minnen2018joint}               & -          & Local[S]        & $>10^3$          & -3.24                        \\
Minnen \emph{et. al.}\cite{minnen2020channel}                   & Even          & -          & 67          & -3.96                               \\
He \emph{et. al.}\cite{he2021checkerboard}           & -         & Local[P]         & 28.2          & -2.91                 \\
Qian \emph{et. al.}\cite{qian2021entroformer}             & -         & Global[S]        &   $>10^3$ &-4.89                                \\
 He \emph{et. al.}\cite{he2022elic}        & Uneven         &-          &  37.2      &
-3.12\\
     He \emph{et. al.}\cite{he2022elic}        & Uneven         & Local[P]          &   156.9       &
-5.71\\  Jiang \emph{et. al.}\cite{jiang2023mlic}        & Uneven         & Local[P]+Global[P]          &  207.4      &
-7.89  \\ \hline
Ours         & Uneven          & Local[S]+Global[S]         &   $>10^3$        & \textbf{-9.38}          \\

Ours          & Uneven         & Local[P]+Global[P]           & \textbf{196.3}         & \textbf{-8.25}                              \\ \hline   VVC &- & - & - &  0.00    \\ \hline\hline

\end{tabular}}

\label{contextmodel}

\end{table}

\section{Conclusion}
We developed a neural image compression model which improves perceptual image quality using a non-isotropic diffusion decoder. This decoder's inductive bias effectively separates frequency components, leading to the creation of high-quality images. Moreover, we introduce an innovative entropy model that optimizes the trade-off between compression efficiency and decoding speed. This entropy model, founded on the Transformer architecture with Laplacian-shaped positional encoding, establishes a strong global spatial context. Our results underscore the efficacy of leveraging diffusion models and advanced entropy modeling to achieve outstanding image compression performance.\\\textbf{Acknowledgement:} This research is based upon work supported by the National Aeronautics and Space Administration
(NASA), via award number 80NSSC21M0322 under the title of \emph{Adaptive and Scalable Data Compression for Deep Space Data Transfer Applications using Deep Learning}.

%We developed a neural image compression model thatimproves perceptual quality of image by using a non-sotropic diffusion decoder. This decoder introduces aninductive bias that effectively discerns between frequencycomponents and enhances the generation of high-qualityimages. In addition, we proposed a novel entropy modelwhich better balances compression ability and runningspeed. The entropy model is constructed based on theTransformer with a Laplacian-shaped positional encoding,enables the construction of a robust global spatial context
{
    \small
    \bibliographystyle{ieeenat_fullname}
    \bibliography{main}
}

\clearpage
\setcounter{page}{1}
\maketitlesupplementary

\section{Denosing Diffusion Models}
Denoising diffusion models are hierarchical latent variable models which generate sample through gradually removing noise from a randomly sampled white noise vector. The training procedure is comprised of two processes: diffusion or forward and  denoising or backward. Diffusion process destroy the clean image and convert it to an approximately pure Gaussian noise during $T$ time steps.  The learnable denoising process then reconstructs the data distribution from white noise by reversing the diffusion process.\\

\noindent \textbf{Diffusion Process:} The diffusion process \cite{ho2020denoising} can be described as a Markov chain, wherein each step of the forward path is defined by a Gaussian transition kernel:
\begin{equation}
q(\bm{z_{t}}|\bm{z_{t-1}}) =\mathcal{N} (\bm{z_t};\alpha_{t|t-1}\bm{z_{t-1}},{\sigma^2_{t|t-1}}\bm{I}),
\label{eq1}
\end{equation}
where ${\alpha_{t|t-1} \in R^+}$ governs the extent to which the previous latent is retained, while ${\sigma_{t|t-1}} \in R^+$ regulates the magnitude of the added noise. The dimension of the latent variables ${\bm{z_1},...,\bm{z_T}}$ is the same as that of the data $\bm{x}$ or $\bm{z_0}$. An important property of the forward process is that any desired step $\bm{z_t}$ can be directly sampled from $\bm{x}$ using a closed-form solution, without needing to compute preceding steps:
\begin{equation}
q(\bm{z_{t}}|\bm{x}) =\mathcal{N} (\bm{z_t};\alpha_{t}\bm{x},{\sigma^2_{t}}\bm{I}),
\label{eq2}
\end{equation}
where ${\alpha_{t|t-1}}= {\alpha_{t}}/{\alpha_{t-1}}$ and $\sigma^2_{t|t-1}=\sigma^2_{t}-\alpha^2_{t|t-1} \sigma^2_{t-1}$. The pre-specified hyperparameters $\alpha_{t}$ typically exhibit a monotonically decreasing pattern from $1$ to $0$, while $\sigma_{t}$ monotonically increases from $0$ to $1$. This pattern leads to a gradual corruption of the input image by Gaussian noise as $t$ increases, resulting in $q(\bm{z_T})=\mathcal{N}(\bm{0},\bm{I})$.\\

\noindent \textbf{Denoising Process:} The true denoising distribution, which is tractable when conditioned on $\bm{x}$ \cite{ho2020denoising}, can be written: 
\begin{equation}
q(\bm{z_{t-1}}|\bm{z_{t}}, \bm{x})=
\mathcal{N} (\bm{z_{t-1}};\bm{\mu_{{t}\to{t-1}}(x,z_t)} ,{{\sigma^2}_{{t}\to{t-1}}\bm{I}}), 
\label{eq3}
\end{equation}
where the distribution parameters can be computed as:
\begin{align}
\begin{split}
& {\sigma_{t \to t-1}}= \sigma_{t|t-1}\sigma_{t-1}/\sigma_t\\& \bm{\mu_{{t}\to{t-1}}}=(\alpha_{t|t-1}\sigma_{t-1}^2/\sigma_{t}^2)\bm{z_t}+(\alpha_{t-1}\sigma_{t|t-1}^2/\sigma_{t}^2)\bm{x}
\label{eq4}
\end{split}
\end{align}
To generate data, the true denoising process can be estimated  by a learned denoising distribution $p_\theta(\bm{z_{t-1}}|\bm{z_t}):=q(\bm{z_{t-1}}|\bm{z_{t}}, \bm{\hat{x}}=\phi_\theta(\bm{z_t},t))$, where $\bm{\hat{x}}$ is  predicted from diffused sample $\bm{z_t}$ using a neural network $\phi_\theta$. Similar to Eq. 4, $p_\theta(\bm{z_{t-1}}|\bm{z_t})$ can be expressed by the approximation $\bm{\hat{x}}$:
\begin{align}
\begin{split}
& p_\theta(\bm{z_{t-1}}|\bm{z_{t}}) =\mathcal{N} (\bm{z_{t-1}};\mu_{\theta}(\bm{z_{t}},\bm{t}), {{\sigma^2}_{{t}\to{t-1}}}\bm{I}),\\& = \mathcal{N} (\bm{z_{t-1}}; \bm{\mu_{{t}\to{t-1}}(\hat{x},z_t)} ,{{{\sigma^2}_{{t}\to{t-1}}}\bm{I}}).
\label{eq5}
\end{split}
\end{align}
\textbf{Training Objective}: The likelihood function $\log p_\theta(\bm{x})$ is challenging to compute directly for training the model. So, during training, its evidence lower bound is maximized ($\text{ELBO}\leq\log p_\theta(\bm{x})$), which can be expressed as:
\begin{multline}
   \text{ELBO} = \bm{E}_{q}[-\overbrace{D_{KL}(q(\bm{z_T}|\bm{x})||p(\bm{z_T}))}^{L_T} \overbrace{+\log p_\theta(\bm{x}|\bm{z_1})}^{L_0}\\ +  \sum_{t=2}^{T}- \overbrace{D_{KL} (q(\bm{z_{t-1}}|\bm{z_{t}}, \bm{x)}||p_\theta(\bm{z_{t-1}}|\bm{z_{t}})}^{L_{t-1}}.
\label{eq6}   
\end{multline}
Within a well-defined  noise scheduling, both $L_0$ and $L_T$ tend to approach approximately $0$ and remain constant. Therefore, for training the diffusion model, it becomes adequate to optimize the $L_{t-1}$ term, which is equivalent to comparing the learnable denoising process with the true denoising distribution. As both of these distributions are Gaussian, the expressions for the KL divergences have closed-form solutions and can be written as follows:
\begin{equation}
 L_{t-1}\propto \bm{E}_{q}[||\bm{\mu_{{t}\to{t-1}}}-\mu_{\theta}(\bm{z_{t}},\bm{t})||^2]=\bm{E}_{q}[||\bm{x}-\bm{\hat{x}}||^2]. 
 \label{eq7}
\end{equation}

In above formulation, the neural network directly predicts $\bm{\hat{x}}$. However, \cite{ho2020denoising} discovered that optimization becomes simpler by predicting Gaussian noise instead. Hence, if we express $\bm{z_t}=\alpha_t\bm{x}+\sigma_t\bm{\epsilon}$, then the neural network $\phi_\theta$ generates $\hat{\bm{\epsilon}}=\phi_\theta(\bm{z_t},t)$, resulting in:
\begin{equation}
\bm{\hat{x}}= (1/{\alpha_t})\bm{z_t}- ({\sigma_t}/{\alpha_t}) \bm{\hat{\epsilon}}.
\label{eq8}
\end{equation}

As demonstrated in \cite{kingma2021variational}, using this specific parameterization, the final loss is obtained as follows:
\begin{equation}
 \bm{E}_{t,\bm{x},\bm{\epsilon}} [||\bm{\epsilon}-\bm{\hat{\epsilon}}||^2]=\bm{E}_{t,\bm{x},\bm{\epsilon}} [||\bm{\epsilon}-\phi_\theta(\alpha_t\bm{x}+\sigma_t\bm{\epsilon},t)||^2.    \label{eq9}
\end{equation}

%we can reparameterize the mean to make theneural network learn the added noise at time step t instead. 

%In this formulation, Ho et al. (2020) observed that optimizing the neural network to predict the Gaussian noise directly, instead of predicting the actual data point ˆx, yields better results. In particular, when  zt is reparametrized as zt = αtx + σtε, the neural network φ generates an estimate e = φ(zt, t), so that:

%As shown inHo et al. [14], we can reparameterize the mean to make the neural network learn the added noise at time step t instead.
%In this way, µθ can be reparameterized as follows: As demonstrated in , employing this specific parametrization leads to a simplified form of Lt.

%The denoising process is also a Markov chain where each step is modeled by a neural network that compute the mean and covariance of a Gaussian conditional distribution:
%p(\bm{z_{t-1}}|\bm{z_{t}}) =\mathcal{N} (\bm{z_{t-1}};\mu_{\theta}(\bm{z_{t}},\bm{t}),{\Sigma_{\theta}}(\bm{z_{t}},\bm{t})),

\section{Additional Details on Blurring Diffusion Model}

\noindent\textbf{Heat Dissipation as Gaussian Diffusion:}
The heat dissipation process or blurring \cite{rissanen2022generative} can be expressed as a type of Gaussian diffusion. First, the marginal distribution of any time step noisy latent $\bm{z_t}$ can be defined as follows:
\begin{equation}
q(\bm{z_{t}}|\bm{x}) =\mathcal{N} (\bm{z_t};\bm{A_{t}}\bm{x},{\sigma^2\bm{I}}),
\label{eq1}
\end{equation}
where $\bm{A_t} = \bm{VD_tV^T}$ represents the dissipation or blurring operation. $\bm{V^T}$ contains orthogonal Discrete Cosine Transform (DCT) basis, while the diagonal matrix $D_t = \exp(-\bm{\Lambda} \tau_t)$ corresponds to the exponentiation of a weighting matrix for the frequencies $\bm{\Lambda}$. $\bm{\Lambda}$ contains squared frequencies $\lambda_{n,m} = -\pi^2(n^2/W^2 + m^2/H^2)$, where $W$ and $H$ are the width and height of the image, and $n \in \{0, \ldots, W-1\}$ and $m \in \{0, \ldots, H-1\}$. According to Eq. \ref{eq1}, any latent state $\bm{z_t}$ is created by introducing a constant level of noise to a progressively blurred data point. When we transform the variables using the following transformations: $\bm{f_t = V^T z_t}$ and $\bm{f_x = V^T x}$, the Gaussian diffusion process can be formulated in frequency space:

\begin{align}
\begin{split}
    &q(\bm{ V^T z_{t}}|\bm{ V^T x}) =\mathcal{N} (\bm{ V^T z_t}; \bm{V^T A_{t}}\bm{x},{\sigma^2\ \bm{V^T \bm{I} V}}) \Leftrightarrow\\
   & q(\bm{f_t}|\bm{ f_x}) =\mathcal{N} (\bm{f_t}; \bm{D_{t}}\bm{f_x},{\sigma^2\  \bm{I}}). 
\end{split}
\end{align}
If we define a vector $\bm{\lambda}$ containing the diagonal elements of $\bm{\Lambda}$, we can express $\bm{d_t}$ as $\exp(-\bm{\lambda} \tau_t)$, which corresponds to the diagonal elements of the matrix $\bm{D_t}$. With this reinterpretation, the diffusion process in frequency space can be written as follows:
\begin{equation}
q(\bm{f_t}|\bm{ f_x}) =\mathcal{N} (\bm{f_t}; \bm{d_{t}} \odot\bm{f_x},{{\sigma^2}\bm{I}}), 
\label{eq3}
\end{equation}
where $\odot$ denotes elementwise vector multiplication. 
Eq. \ref{eq3} shows that the marginal distribution of $\bm{f_t}$ can be decomposed into individual scalar elements $\bm{f_t^{(i)}}$. Likewise, the learnable inverse heat dissipation model $p_\theta(\bm{f_{t-1}}|\bm{f_t})$ can also be decomposed in a fully factorized manner. As a result, we have the option to describe the heat dissipation process and its inverse using scalar representations for each dimension $i$:
\begin{align}
\begin{split}
& q({f_t^{(i)}}|{f_x^{(i)}}) =\mathcal{N} ({f_t^{(i)}}; d_{t}^{(i)} u_x^{(i)},{\sigma^2}) \Leftrightarrow\\
& f_t^{(i)}= d_{t}^{(i)} f_x^{(i)}+ {\sigma}{\epsilon}, \text{with} {\epsilon} \sim \mathcal{N}(0,1).
\label{eq4}
\end{split}
\end{align}
Eq. \ref{eq4} can be identified as a particular case of the standard Gaussian diffusion process that operates in frequency space, i.e., $f_t^{(i)} = {\alpha}_t f_x^{(i)} + {\sigma}_t{\epsilon}$, where ${\alpha}_t = d_{t}^{(i)}$ and ${\sigma}_t = {\sigma}$. What distinguishes this type of diffusion process from the standard one is the utilization of distinct noise schedules, denoted as $\alpha_t$ and $\sigma_t$, for each scalar element of the latent variable $\bm{f_t}$. In other words, the noise applied in this process exhibits non-isotropic characteristics. It's worth noting that while the marginal variance $\sigma$ is shared across all scalar elements $f_t^{(i)}$, the specific noise schedules provide individual adjustments for each element.

In heat dissipation models, the Markov process $q(\bm{f_t}|\bm{f_{t-1}})$ can be defined, corresponding to their chosen marginal distribution $q(\bm{f_t}|\bm{f_{x}})$. By establishing an equivalence with Gaussian diffusion, this process can be effectively described using the following formulation:
\begin{align}
\begin{split}
& q(\bm{f_t}|\bm{f_{t-1}}) =\mathcal{N} (\bm{{f_t}}; \bm{\alpha_{t|t-1}} \bm{f_{t-1}}, \bm{{\sigma^2_{t|t-1}}\bm{I}}),\\
&\text{where }  \bm{\alpha_{t}}=\bm{d_t},\sigma_t^{(i)}=\sigma \Rightarrow \bm{\alpha_{t|t-1}}\bm{=\frac{d_t}{d_{t-1}}},\\& \Rightarrow \bm{{\sigma^2_{t|t-1}}}= (1-(\bm{\frac{d_t}{d_{t-1}}})^2)\sigma^2.
\end{split}
\end{align}

When $\bm{d_t}$ is designed to have smaller values for higher frequencies, $\bm{\sigma_{t|{t-1}}}$ will introduce greater noise to the higher frequencies at each timestep. This results in the heat dissipation model erasing information from those frequencies more rapidly compared to the standard diffusion process.

\noindent \textbf{Inverse Heat Dissipation:} Similar to the standard diffusion model \cite{ho2020denoising}, the analytical expression for the true inverse heat dissipation process is obtained and can be written as follows:

\begin{equation}
q(\bm{f_{t-1}}|\bm{f_{t}}, \bm{f_{x}})=
\mathcal{N} (\bm{f_{t-1}}; \bm{\mu_{{t}\to{t-1}}} ,\bm{{{\sigma^2}_{{t}\to{t-1}}}}\bm{I}), 
\end{equation}

\noindent where:
\begin{align}
\begin{split}
& q(\bm{f_{t-1}}|\bm{f_{t}}, \bm{f_{x}}) \propto q(\bm{f_{t-1}}|\bm{f_{x}}) 
 q(\bm{f_{t}}|\bm{f_{t-1}},\bm{f_{x}})=\\
& q(\bm{f_{t-1}}|\bm{f_{x}}) q(\bm{f_{t}}|\bm{f_{t-1}}) \Rightarrow \bm{{\sigma_{t \to t-1}}= \sigma_{t|t-1}\sigma_{t-1}/\sigma_t},\\&
\bm{\mu_{{t}\to{t-1}}}= (\bm{\alpha_{t|t-1}\sigma_{t-1}^2}/\bm{\sigma_{t}^2})\bm{f_t}+(\bm{\alpha_{t-1}\sigma_{t|t-1}^2}/\bm{\sigma_{t}^2})\bm{f_x}
\end{split}
\end{align}
As discussed, the true denoising process can be approximated using a learned denoising distribution, $p_\theta(\bm{f_{t_1}|\bm{f_t}})$.
\section{Algorithms}

Algorithms 1 and 2 summarize the training and decoding procedures of our neural codec.

\begin{algorithm}
\caption{Training Neural Codec}\label{alg:cap}
\begin{algorithmic}
\State Sample $\bm{x}\sim$dataset
\Repeat
\State Sample $t\sim\mathcal{U}(0,T)$
\State Sample $\bm{\epsilon}\sim\mathcal{N}(\bm{0},\bm{I})$
\State $\bm{z_t}=\bm{V}\bm{\alpha_t}\bm{V^T}\bm{x}+\bm{V}\bm{\sigma_t}\bm{V^T}\bm{\epsilon}$
\State $\bm{\Tilde{y}}\sim \mathcal{U}(En_\zeta(\bm{x})-0.5,En_\zeta(\bm{x})+0.5)$
\State $\bm{\hat{x}_t}=\bm{V}({1}/{{\bm{\alpha_{t}}}})(\bm{V^T z_t}-{{\bm{\sigma_{t}}}} {\bm{V^T} \phi_\theta(\bm{z_t},t,\bm{\tilde y}}))$
\State $L_{Dif}=||\bm{\epsilon}-\phi_\theta(\bm{z_t},t,\bm{\tilde y})||^2$
\State $L_T=(1-\beta)L_{Dif}+\beta d_{\text{LPIPS}}(\bm{x},\bm{\hat{x}_t})
-\lambda \log p_{\bm{\tilde{y}}}(\bm{\tilde{y}})$
\State $(\zeta,\theta)=(\zeta,\theta)-\eta\nabla_{\zeta,\theta}L_T$ ($\eta:$ Learning Rate)
\Until{ converged}
\end{algorithmic}
\end{algorithm}

\begin{algorithm}
\caption{Decoding Compressed File}\label{alg:cap}
\begin{algorithmic}
\State $\bm{\hat{y}}\Leftarrow$ Entropy decoded binary file using entropy model $p_{\hat{\bm{y}}}(\hat{\bm{y}})$ 
\State Sample $\bm{z_T}\sim\mathcal{N}(\bm{0},\bm{I})$
  \For{$t = T,..., 1$ }
  \State $\bm{f_t}=\bm{V^T z_t}$ and $\bm{{f_{\hat{\epsilon}}}}=\bm{V^T}\phi_\theta(\bm{z_t},t,\bm{\hat y})$
  \State$\bm{{\sigma_{t \to t-1}}= \sigma_{t|t-1}\sigma_{t-1}/\sigma_t}$
  \State $\bm{\hat{\mu}_{t \to t-1}}=\frac {\bm{\alpha_{t|t-1}\sigma_{t-1}^2}}{\bm{\sigma_{t}^2}} \bm{f_t}+\frac{\bm{\sigma_{t|t-1}^2}}{\bm{\alpha_{t|{t-1}}}{\bm{\sigma_t}^2}}(\bm{f_t}-\bm{\sigma_t}\bm{{f_{\hat{\epsilon}}}})$
  \State $ \bm{z_{t-1}}\leftarrow \bm{V}(\bm{\hat{\mu}_{t \to t-1}}+\bm{\sigma_{t \to t-1}}\bm{{f_{\hat{\epsilon}}}})$
  \EndFor \State \textbf{Return} $\bm{\hat{x}}=\bm{z_0}$
\end{algorithmic}
\end{algorithm}

\section{Architecture of Diffusion-based Decoder}

\begin{figure*}
  \centering
  \includegraphics[width=0.86\linewidth]{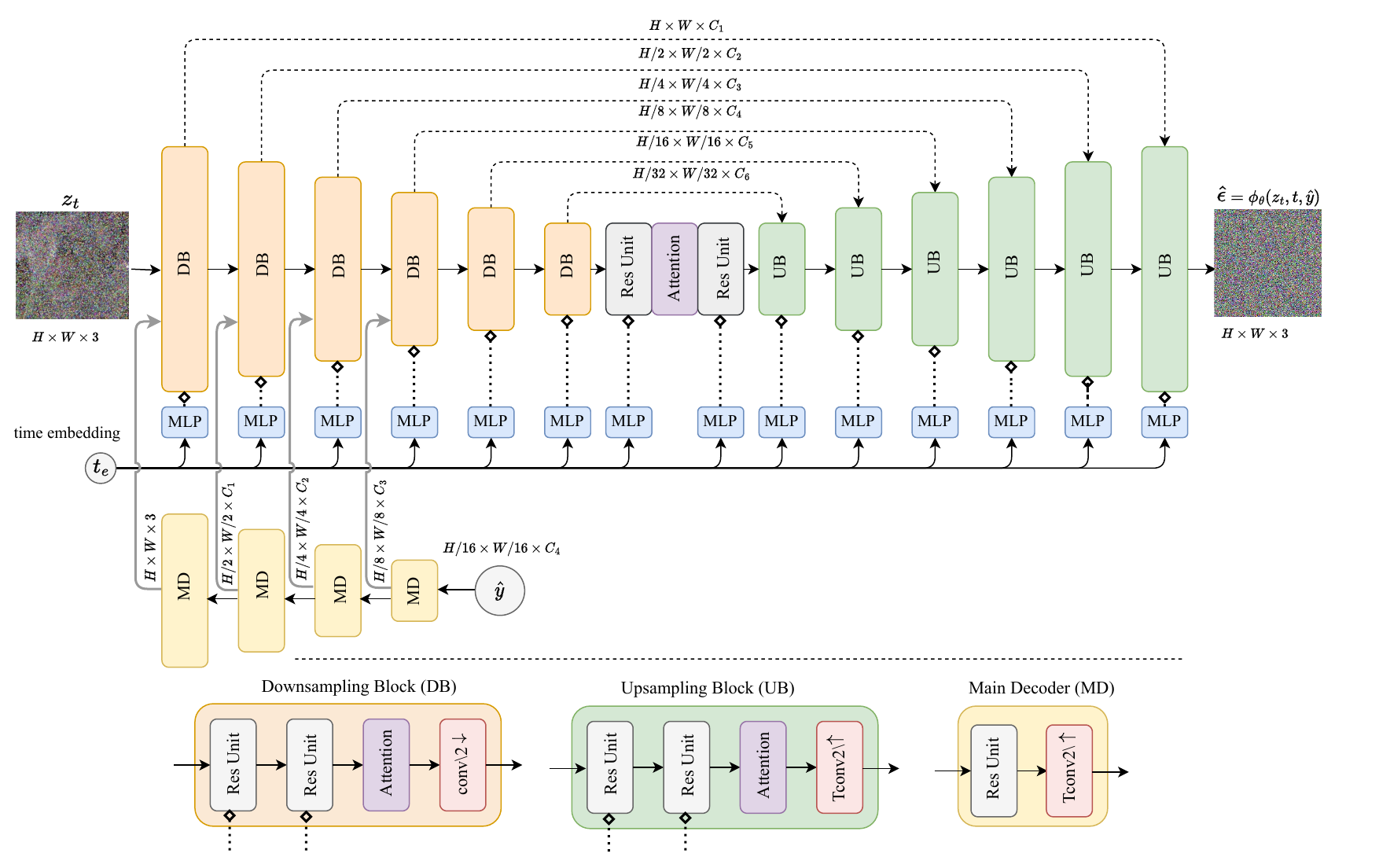}
  \caption{ Architect of diffusion-based decoder. $W$ and $H$ correspond to the width and height of the input image, respectively. }
  \label{fig:1}
\end{figure*}

Fig. \ref{fig:1} illustrates our diffusion-based decoder design, employing a U-Net architecture for the diffusion model \cite{ho2020denoising}, incorporating ResNet blocks and self-attention modules. We've employed six units for both encoding and decoding within the U-Net framework. In the encoding pathway, the channel dimension is determined from the set $\{C_1=64,C_2=128,C_3=192,C_4=256,C_5=320,C_6=384\}$. The decoding process mirrors the encoding process in reverse. The main decoder (MD) comprises ResNet blocks and transposed convolutions, which serve to upscale the quantized latent representation $\bm{\hat{y}}$ to match the spatial dimensions of the inputs from the initial $4$ U-Net encoding units. This setup enables us to introduce conditioning by concatenating the output of the main decoder layers with the input from the corresponding U-Net layer. 

The time step $t$ is initially linearly embedded into a vector with a dimension of $64$. Subsequently, the resulting time embedding $t_e$ is further processed through MLP layers, which are responsible for expanding it to align with the channel size of the corresponding DB/UB layers.

\begin{figure*}[tp]
    \centering
    \includegraphics[width=1\linewidth]{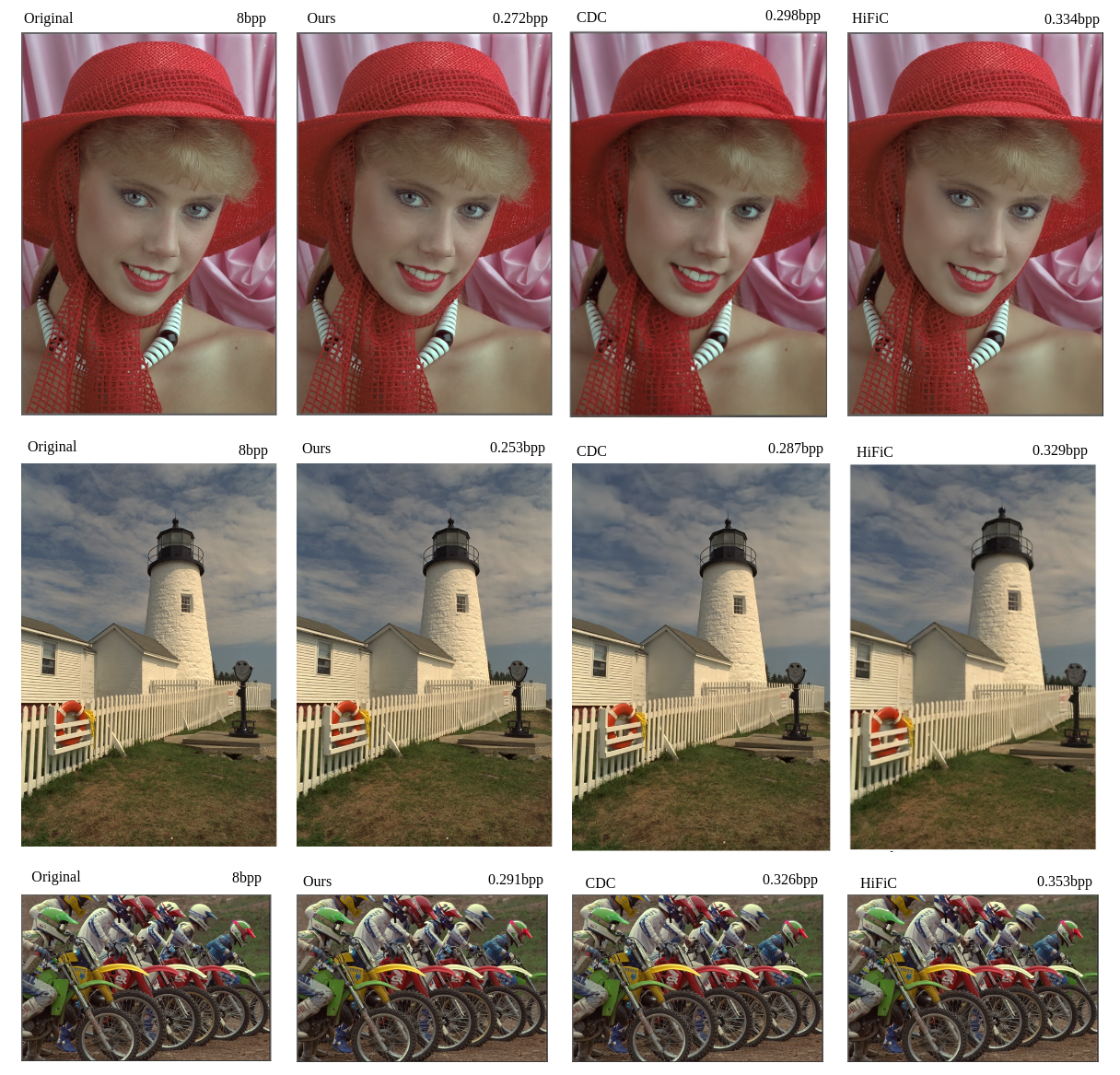}
    \caption{ Additional Visual comparison of our method to the HiFiC and CDC models.}
    \label{fig:2}
\end{figure*}

\section{Additional Qualitative Comparisons }
As shown in Fig. \ref{fig:2}, our model tends to generate fewer artifacts and is capable of decoding images with greater realism compared to both the HiFiC \cite{mentzer2020high} and CDC \cite{yang2022lossy} networks, even when using a significantly lower bit-rate.

\end{document}